\documentclass[electronic]{vgtc}             

\ifpdf
  \pdfoutput=1\relax                   
  \usepackage{graphicx}                
  \DeclareGraphicsExtensions{.pdf,.png,.jpg,.jpeg} 
\else
  \usepackage{graphicx}                
  \DeclareGraphicsExtensions{.eps}     
\fi%

\graphicspath{{figures/}{pictures/}{images/}{./}} 

\usepackage{microtype}                 
\PassOptionsToPackage{warn}{textcomp}  
\usepackage{textcomp}                  
\usepackage{mathptmx}                  
\usepackage{times}                     
\usepackage{cite}                      
\usepackage{tabu}                      
\usepackage{booktabs}                  

\usepackage{eurosym}
\usepackage{multirow}
\usepackage[table]{xcolor}
\usepackage{amsthm}
\usepackage{subfig}
\usepackage{float}

\onlineid{0}

\vgtccategory{Research}

\vgtcinsertpkg

\title{QuViS - The Question of Visual Site Selection}

\author{Sebastian Baumbach\thanks{e-mail: sebastian.baumbach@dfki.de}
\and Jahanzeb Khan
\and Sheraz Ahmed
\and Andreas Dengel
}
\affiliation{\scriptsize German Research Center for Artificial Intelligence (DFKI), Kaiserslautern, Germany \\
\scriptsize University of Kaiserslautern, Kaiserslautern, Germany}

\teaser{
	\centering
  \fbox{\includegraphics[width=0.75\columnwidth]{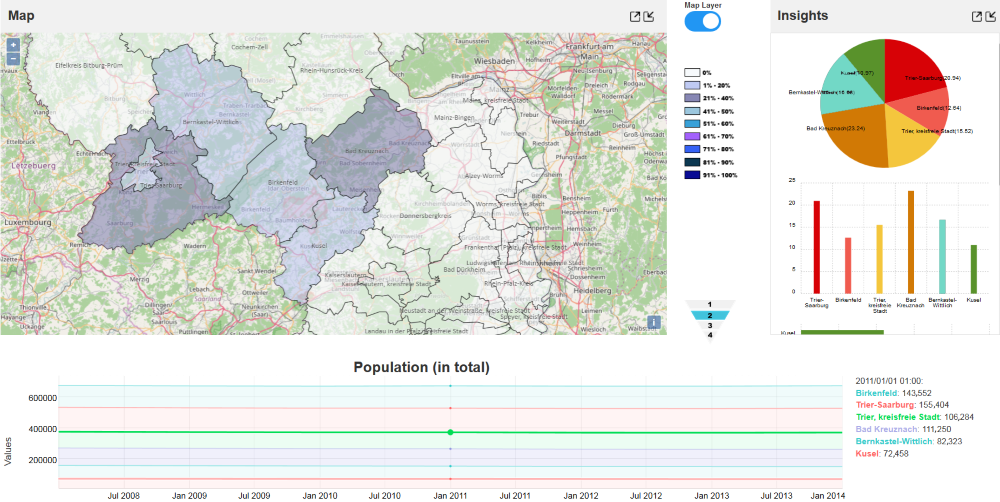}}
  \caption{\textit{QuViS} supports decision-makers in companies in performing exploratory visual data analysis on geospatial-temporal data to select suitable sites.}
 }

\abstract{
This paper present \textit{QuViS}, which is an interactive platform for visualization and exploratory data analysis of site selection. The aim of \textit{QuViS} is to support decision makers and experts during the process of site selection. In addition to visualization engine for exploratory analysis, \textit{QuViS} is also integrated with our automatic site selection method (QuIS), which recommend different sites automatically based on the selected location factors by economists and experts. To show the potential and highlight the visualization and exploration capabilities of \textit{QuViS}, a case study on $1,556$ German supermarket site selection is performed. The real publicly available dataset contains $450$ location factors for all $11,162$ multiplicities in Germany, covering the last $10-15$ years. Case study results shows that \textit{QuViS} provides an easy and intuitive way for exploratory analysis of geospatial multidimensional data. 
} 

\CCScatlist{ 
  \CCScat{H.3.3}{Information Search and Retrieval}{Information Search and Retrieval}{Search process};
  \CCScat{H.5.2}{Information Interfaces and Presentation}{Graphical user interfaces (GUI)}{User Interfaces}
}

\begin{document}


\firstsection{Introduction}
\maketitle

Selecting a facility location is an important investment decision for every company that wants to establish a successful business~\cite{Strotmann2007, Bhatnagar2005443}. Enterprises encounter such decision problems at least once in their lifetime. Numerous geographical, social, political, economic, and socio-economic factors for each site have to be taken into account by economists and compared to the specific requirements of a company~\cite{blair1987major}. After a site is chosen and the business is operational at the selected site, negative influences of this site can hardly be compensated~\cite{woratschek2000, glatte2015location}.
\newline
In the modern era of globalization and information explosion, countless information sources about locations are widely available. This includes geographic data (Google Maps or OpenStreetMap), economic, and social attributes of inhabitants (for instance, offered by the Federal Statistical Office of Germany as open data). In addition to open data sources, there exist many companies focusing on the creation of complex and commercial datasets, ranging from the average buying power of municipalities to consumer behavior of citizens. As a consequence, multidimensional and heterogeneous datasets are available, containing spatial and temporal information that is of high relevance for tasks like site selection. This exponential growth of decisive data, however, makes it increasingly difficult for experts to assess all information, which are relevant for selection the optimal sites. 
\newline
The economists usually perform exploratory data analysis for the purpose of site selection. This means that they analyze the data by visualizing it using different tools. The existing approaches for visualization of spatial-temporal data are not suitable for selection sites as they mostly focus on point-wise data, such as tweets, or trajectories of people. The main problem with point-wise data is that the spatial entities are moving over time. Second, these point-wise data representing abstract points in space and thus, have to be projected and visualized on a map. In contrast, the objects of analysis in site selection \textit{are} sites existing in the real world, which are characterized by geo-referenced location factors over time. Decision-makers have to perform exploratory analysis by exploring sites and comparing their descriptive factors with the companies' requirement. More important, locations are selected top -- down (i.e. from state to region) in practice. 
Consequently, some regions are excluded subjectively by experts and the selection is commonly based on subjective criteria \cite{mcmillan1965manufacturers, luder1983unternehmerische, Zelenovic2003}. However, no visual representation in literature utilizes the administrative hierarchy of nations in the way that is necessary in the site selection process.
\newline
The aim of this paper is to design a system for exploratory visual data analysis to support decision-makers of companies in site selection. In our previous work, we proposed a data driven quantitative model for site selection, which contains all locations as well as their descriptive location factors \cite{baumbach2017QuIS}. This model describes sites by their attributes, e.g. \textit{purchasing power}, \textit{number of inhabitants} or \textit{proximity to suppliers}. Furthermore, we proposed \textit{QuIS} as a methods for data-driven site selection, which analyzes location factors and company requirements in order to recommend suitable sites. 
\newline
We propose \textit{QuViS}, which is especially designed for interactive data exploration and visualization for site selection. \textit{QuViS} follows Shneiderman’s  "visual information-seeking mantra: overview first, zoom and filter, then details-on-demand" \cite{shneiderman1996eyes}. Our proposed design for visual site selection provides users the ability to \textit{browse} through all sites, \textit{select} one or multiple sites in order to \textit{inspect} their location factors and \textit{compare} sites to find the optimal one. All these operations are supported by data \textit{filtering} capabilities along the time, space and feature (here, location factor) dimension. More important, the administrative hierarchy is incorporated into the data exploration as sites are selected top -- bottom. Sites and location factors are \textit{aggregated} along this administrative hierarchy, which enables users to drill down by expanding along this hierarchy. The applied data model supports data aggregation along the administrative hierarchy by itself. Finally, on the municipality respective city level, user can assess each individual site in detail to ultimately select the most suitable one.
\newline
In particular, this paper made the following contributions:

\begin{itemize}
	\item A characterization of the application domain \textit{site selection} with respect to a detailed task description experts need to perform for selecting sites.
	\item The design decisions we made to construct a visual representation which supports the application scenario of site selection.
	\item The visual encoding and the interaction mechanisms we have chosen based on the design decisions to support users in the exploratory tasks behind site selection.
	\item A case study of German supermarkets highlighting the possibilities our proposed approach.
\end{itemize}


\section{Site Selection: An Overview}
\label{sec:Background}

The \textit{site} or \textit{location} of a company is the geographic place where a company does business~\cite{luder1983unternehmerische}. The location characteristics, which are relevant for the operating performance, are called \textit{location factors}~\cite{weber1909urber}. While the economic viewpoint examines the global distribution of companies from a broader perspective, the  intra-company viewpoint deals with the spatial arrangement of resources within buildings, which is also known as layout planning. In this paper, the focus is on the business perspective of site respective location selection rather than on the economic or intra-company point of view. 
\newline
The rest of this section gives an introduction into the current state of research on locational datasets (containing both a list of sites and location factors) and site analysis in science and industry. We point interested readers to our previous work \cite{baumbach2017QuIS} for a detailed review of site selection methods as well as our proposed recommender system for sites named \textit{QuIS}.

\subsection{Location Factors}
\label{subsec:Background_location_factors}

Site selection is mainly done by analyzing and comparing the location factors of different places. These factors are the properties that are influential towards a company's goal achievement~\cite{liebmann1969grundlagen}. The important location factors are usually selected by a company based on their own demands and each company might possess different requirement. \autoref{tab:LocationFactors} shows the most common location factors grouped into 10 major categories~\cite{badri2007dimensions}. 

\begin{table}[!ht]
	\def\arraystretch{1.25}
	\caption{Categorized Location Factors, a Selection}
	\label{tab:LocationFactors}
	\centering
	\resizebox{1\columnwidth}{!}{
		\begin{tabular}{l|p{7.5cm}}
			\textbf{Category} & \textbf{Location Factor} \\ 
			\hline
			Transportation           & Highway facilities, railroad facilities, waterway transportation, airway facilities, trucking services, shipping cost of raw material, cost of finished goods transportation, shipping cost of raw material, warehousing and storage facilities.  \\ 
			\hline
			Labor                    & Low cost labor, attitude of workers, managerial labor, skilled labor and wage rates, unskilled labor, unions, and educational level of labor. \\ 
			\hline
			Raw Materials            & Proximity to supplies, availability of raw materials, nearness to component parts, availability of storage facilities for raw materials and components, and location of suppliers. \\
			\hline
			Markets                  & Existing consumer market, existing producer market, potential consumer market, anticipation of growth of markets, marketing services, favorable competitive position, population trends, location of competitors, future expansion opportunities, and size of market. \\ 
			\hline
			Industrial Site          & Accessibility of land, cost of industrial land, developed industrial park, space for future expansion, availability of lending institution, closeness to other industries. \\ 
			\hline
			Utilities                & Attitude of utility agents, water supply, wastewater, cost and quality, disposable facilities of industrial waste, availability of fuels, cost of fuels, availability of electric power, and availability of gas. \\ 
			\hline
			Government Attitude      & Building ordinances, zoning codes, compensation laws, insurance laws, safety inspections and stream pollution laws. \\ 
			\hline
			Tax Structure            & Industrial property tax rates, state corporate tax structure, tax free operations, and state sales tax. \\ 
			\hline
			Climate \& Ecology       & Amount snow fall, percent rain fall, living conditions, relative humidity, monthly average temperature, air pollution, and ecology. 
		\end{tabular}
	}
\end{table}

Location factors can be further divided into hard and soft factors. The first group is quantifiable and has direct impact on the economic viability of locations as they directly influence the cost and revenue calculations. This includes, among others, the number of skilled workers and their education, the proximity to suppliers, and tax rates. The second group contains qualifiable factors, which cannot be integrated into the calculations of the company due to their descriptive and non-numerical nature. This comprises, for instance, political conditions, prestige of sites, or leisure programs. However, the later group is becoming more and more important for site analysis~\cite{blair1987major}.
\newline
These criteria are highly dependent on the industry and the specific company. For retail stores, the close proximity (purchasing power, public transport, or prestige) is of crucial importance. In comparison, the availability of labor, infrastructure, and tax considerations are of high interest for factories. Sites for the production of hazardous substances require the lack of residential and other industrial facilities in their close surrounding.

\subsection{Site Selection in Science and Industry}
\label{subsec:Background_site_selection_methods}

Many approaches have been proposed how a site can be selected based on location factors of sites and the given requirements of companies. In the area of site selection, research work traces back nearly a century~\cite{lehr1885mathematische, mcmillan1965manufacturers, weber1909urber}. In fact, there is no fixed procedure. However, few guidelines and operating procedures exist.
\newline
In literature and practice, the site selection process is often divided into multiple phases where the regional focus is reduced in each phase. Zelenovic split it up into a macro and micro selection \cite{Zelenovic2003}. The first phase addresses the issue of finding the right state while the second one looks for a specific site within the previously chosen state. Bankhofer divided the whole process into four phases of selecting continent, country, municipality and then final location in that order \cite{Bankhofer2001}. 
\newline
After a list of alternatives is preselected, the final decision about the new site has to be made. Woratschek and Pastowski studied different methods in economics used for this step, such as \textit{Checklist Methods} \cite{woratschek2004dienstleistungsmanagement}. This method is a very basic way to assist in the selection process. Relevant location factors for a company get listed and weighted by experts for different sites. These weights are the degree how good a location fulfills a given requirement. Table \ref{tab:Checklist} shows an example of a checklist for site selection.

\begin{table}[!ht]
	\def\arraystretch{1.25}
	\caption{Checklist Method, Example.}
	\label{tab:Checklist}
	\centering
	\resizebox{1\columnwidth}{!}{
		\begin{tabular}{l|c|c|c}
			\textbf{Location Factors}	& \textbf{Location 1} & \textbf{Location 2} & \textbf{Location 3} \\
			\hline
			Availability of Resources & + & + & o \\
			\hline
			Income Structure & + & - & + \\
			\hline
			Consumer Structure & - & o & o \\
			\hline
			Infrastructure & + & + & + \\
			\hline
			Taxes & o & o & + \\
		\end{tabular}
	}
\end{table}

\subsection{Analysis Tasks for Site Selection}
\label{subsec:Background_Conclusion}

As a consequence, companies do not only have to take all necessary location factors into account while searching for a new site. They also assess and compare all available sites weather they fulfill their specific requirements in new sites -- independent of whether the site is manually selected or automatically recommended. These requirements, however, differ from company to company. This ultimately means that the decision-maker in a company has to analyze and evaluate an enormous amount of data in order to select a suitable site. 

 Therefore, the visualization system has to enable the user to interactively browse through the data, explore it visually, and gives him the ability to efficiently compare sites to each other. However, this exploration and interaction procedure has to maintain the conventional selection process top -- down. Summarizing these findings, we identified the following tasks:

\begin{enumerate}
	\item Task: Browsing through the geographic dimension in order to visually assess the spatial properties of sites, such as the distance between sites, the proximity of a site, or the geographic location in general.
	\item Task: Drill-down or roll-up through the administrative hierarchy, i.e., changing the focus point of the analysis from  states over countries to municipalities respective cities.
	\item Task: Exploring the location factors of sites.
	\item Task: Comparing sites to each other where this comparison is based on the users preselected location factors.
	\item Task: Assessing the change of location factors over time in order to judge the future as these factors are constantly changing by nature. 
\end{enumerate}

Nowadays, the last Task (no. 5) becomes more and more important for site selection in a rapidly changing world. Not only the status quo with its current value for a given location factor is relevant, but also its development over time. For example, if two cities have the same economic strength right now, but one of the cities was a former economical center and is on the decline, whereas the second city is developing magnificently. A decision-maker will most likely chose the second city as this site will most likely outpace the other one in the future.

\section{Design Decision for QuViS}
\label{sec:Design_Decision}

The ultimate goal in site selection is to identify optimal sites where the sites' location factors fulfill the given criteria of the company. Thus, this task falls into the category of data exploration where new, previously unknown information about spatial-temporal sites are revealed \cite{maceachren1994visualization}. In accord with this exploratory perspective, we made our design decision based on the tasks identified in \autoref{subsec:Background_Conclusion} and the extensive work of Andrienko et al. \cite{andrienko2003exploratory} and Kjellin et al. \cite{kjellin2010evaluating}. As these authors state, the tasks as well as the data likewise are constraining the appropriate visualizations that are supportive and efficient for the given usage. 
\newline
In general, our design follows the Visual Information-Seeking Mantra \cite{shneiderman1996eyes} of "Overview first, zoom and filter, then details on-demand". The visual design provides overviews over all sites and gives users the ability to filter sites according to the location factors. In addition, users are able to drill down by expanding along the dimensions, which are the administrative hierarchy and the location factors in this case. On the level of the highest granularity (which are currently municipalities), the exploratory operations of interest for site selection are \textit{identify} and \textit{compare}. Ultimately, the focus lies on sites within the multidimensional data (no data transformation is required) and the interactive exploration of location factors of sites.

\subsection{2D or 3D - That Is the Question}
\label{subsec:Design_2D_vs_3D}

One of the first question while designing visualization systems for multidimensional data is the choice of 2D or 3D representations, which is highly debatable and complex in answering appropriately. Many studies have theoretically assessed or empirically testified the advantages and disadvantages of 2D and 3D representation for gospatial-temporal data. For a thorough review of this field, we point interested readers to the extensive work of Andrienko et al. \cite{andrienko2003exploratory, andrienko2006exploratory, andrienko2010space} as well as MacEachren and Kraak \cite{maceachren1995maps, maceachren2001research}.
\newline
Previous work suggested that no visualization technique is always advantageous in comparison to others, but 2D and 3D representation are counterparts and complement each other \cite{hicks2003comparison, kjellin2010evaluating}. For geospatial-temporal datasets, however, user study conducted by Kaya et al. \cite{kaya20143d} as well as Kjellin et al. \cite{kjellin2010evaluating} propose that 3D visualizations can be advantageous over 2D representation. These authors state that it highly depends on the datasets to be analyzed, in conjunction with the specific tasks a user have to perform on this particular dataset. 
In site selection, however, users identify, retrieve and compare sites to each other, given a small set of location factors. As described in \autoref{subsec:Background_site_selection_methods}, they perform the data analysis from top to bottom and thus, they typically do not have to make analysis based on the whole datasets. As a consequence, 2D representation of data tends to be beneficial for the application of site selection. So far, however, no user studies in this application scenario have been conduct yet so that no clear evidence is given.
\newline
Ultimately, 2D representation is chosen as all of the above-mentioned aspects indicate that this visual encoding is more convenient for site selection. There is no justification that the benefits of 3D representation outweigh the costs \cite{munzner2008process}.

\subsection{Data Classification: Geo-referenced Location Factors over Time}
\label{subsec:Design_Data_Classification}

There are two kinds of data necessary to perform a site selection appropriately: sites and location factors. Both together form the spatial-temporal data, which can be described as geo-referenced location factors over time. However, the application scenario of site selection defines more implicit and explicit properties of the data.

Sites are geospatial\footnote{Geospatial data is defined in the ISO/TC 211 series of standards as data "with implicit or explicit reference to a location relative to the Earth". Source: http://www.isotc211.org/Terminology.htm} objects which are characterized by their geospatial attributes. They are basically defined by the spatial dimension. First, these attributes explicitly include 2D space coordinates (where is the site located), but also geospatial data, such as shapes, geometries, area, and centroid coordinates. Secondly, there are implicit information given, which imply the administrative hierarchy (i.e., a particular municipality belongs to a specific country, which is part of state) as well as inferable spatial properties (e.g. adjacent sites, distances between sites, or general location information, such as located at the sea).

In contrast to geometry-based spatial datasets where the data \textit{represents} points in space (such as geo-referenced tweets or GPS coordinates of people), our data \textit{are}  spatial objects (sites), which are described in detail by location factors (i.e., the cities' population count). This means for the visual encoding, that no spatial data (like points) has to be drawn on the map in addition to administrative division and borders. In fact, the existing sites (countries or districts) need to encode the embedded information.

Besides the given information and in contrast to other typologies of spatial-temporal data, the "spatial extension of objects" involved in this application needs a deeper consideration \cite{Kisilevich2010}. Here, we do not deal with the popular case of point-wise objects, but with areas representing sites. However, the sites are associated with a fixed location and thus, are not moving or changing their spatial attributes over time. Andrienko et al. classified these kind of objects as \textit{static spatial objects} as their "spatial position is constant" and it "exists during the whole time period" \cite{andrienko2011event}.

Location factors introduce the feature dimension (sometimes called non-spatial attributes in literature) and the temporal dimension (since attributes of sites are changing by nature). With respect to the data classification, we have the full history of the location factor and not only the latest values, thus forming a time series. 

\subsection{Exploration vs. Analysis}
\label{subsec:Design_Exploration_vs_Analysis}

As Wang et al. stated in their work \cite{wang2017gaussian}, "the fundamental difficulty" in the visualization of large datasets are defined in two contradicting requirements. First, there is the demand of analyzing the data in various forms and aspects in order to perform comprehensive exploratory analysis. Second, high quality user experiences require an ease while applying the "exploratory hypothesis cycle", which involves a sequence of queries, their execution, result assessment, and finally refining the original query. However, the first requirement demands for powerful analysis functionalities, whereas the second requires computationally efficient implementation. Deriving design decision from these two requirements ends up in contracting objectives \cite{wang2017gaussian}.

For site selection, however, we can arrange the users' wish of a powerful analysis fulfilling their needs with an efficient data handling. The operations, our users like to perform on the data, are narrowed in the application scenario of site selection. This means that we only have to give users the possible to mine the data according to the \textit{tasks} described in \autoref{subsec:Design_Operational_Tasks}. Consequently, the users are satisfied with these possibilities as they can explore sites according to their needs.

Additionally, we do not need to incorporate powerful, but complex analysis algorithms as provided, among others, by Nancubes \cite{lins2013nanocubes} or Gaussian cubes \cite{wang2017gaussian}. These techniques are based on the classic OLAP data cube \cite{gray1997data} and carefully pre-compute aggregations across different subsets of the data to support diversified aggregation operation. In site selection, however, aggregations over arbitrary regions and time spans are not required. We can exploit the implicit and explicit properties of the data (see \autoref{subsec:Design_Data_Classification}), which offers an exploratory analysis by itself. Aggregations are provided along the administrative hierarchy by nature. The population in a specific country is the sum of all of its municipalities. Furthermore, the administrative hierarchy also \textit{slice and dice} the data into states, countries, districts, and municipalities respective cities. Each of these spatial objects have the corresponding location factors assigned. Ultimately, OLAP-based visual analysis, according to the users' needs in site selection, is naturally provided by the data model.

\subsection{Operational Tasks: The Question of Where}
\label{subsec:Design_Operational_Tasks}

In literature, the potential information needs, which can be satisfied by analyzing data, are defined by the data's components~\cite{bertin1983semiology}. For multidimensional data, as given by our geospatial-temporal data, Peuquet \cite{peuquet1994s} distinguished three components. Applied on the corresponding dimensions as stated in \autopageref{subsec:Design_Data_Classification}, we identified: space (\textit{where}), time (\textit{when}) and location factors (\textit{what}). Thus, we can ask three \textit{question type} in site selection, adopted from the work of Bertin \cite{bertin1983semiology}:

\begin{itemize}
	\item \textit{$when + where \rightarrow what$:} Describe the location factors of a site defined by a given location at a given time. For example: \textit{How does Munich currently look like?}
	\item \textit{$when + what \rightarrow where$:} Describe sites at a given time, which have the given location factors fulfilled. For example: \textit{What are the sites with the highest purchasing power index right now?}
	\item \textit{$where + what \rightarrow when$:} Describe the time when a given site had fulfilled a given location factor. For example: \textit{When had Berlin the lowest unemployment rate?}
\end{itemize}

In site selection, however, the question type \textit{$when + what \rightarrow where$} is the \textit{search target} of interest since the focus of analysis mainly lies on sites. Schematically, sites need to be discovered for the given information of \textit{when} and \textit{what}.

The chosen \textit{search target} defines the way our geospatial-temporal data cube is pivoted\footnote{Rotating a data cube is called pivoting in OLAP.}. The next step in designing a visualization system is to specify the \textit{cognitive operations} users can perform on the data. Cognitive operations are defined as possible tasks users can perform on the data \cite{andrienko2003exploratory}. \autoref{tab:CognitiveOperations} shows a list of different cognitive operations while focusing geographical systems, which was defined by Koua et al. \cite{koua2006evaluating}, based on the work of Wehrend and Lewis \cite{wehrend1990problem}.

\begin{table}[!ht]
	\def\arraystretch{1.25}
	\caption{Cognitive Operations for Geographical Systems.}
	\label{tab:CognitiveOperations}
	\centering
	\resizebox{1\columnwidth}{!}{
		\begin{tabular}{r|l}
			Identify & Characteristics of an object \\
			\hline
			Locate & Absolute or relative position \\
			\hline
			Distinguish & Recognize as the same or different \\
			 \hline
			Categorize & Classify according to some property (e.g., color, position, or shape) \\
			\hline
			Cluster & Group same or related objects together \\
			\hline
			Distribution & Describe the overall pattern \\
			 \hline
			Rank & Order objects of like types \\
			 \hline
			Compare & Evaluate different objects with each other \\
			\hline
			Associate & Join in a relationship \\
			\hline
			Correlate & A direct connection \\
			\hline
		\end{tabular}
	}
\end{table}

The tasks in site selection, as described in \autoref{subsec:Background_Conclusion}, can be used to identify cognitive operations users in site selection have to perform on the data. \textbf{Task 1} (Browsing through Space) naturally implies the operations \textit{Locate} and \textit{Categorize}. \textbf{Task 3} (Exploring Location Factors) requires the operation \textit{Identify}. \textbf{Task 4} (Comparing Sites) needs the operator \textit{Compare} as a matter of course, also implies the operation \textit{Rank} as a comparison of all sites results in a ranked list of these sites. \textbf{Task 5} (Assessing the Time) also implies the operations \textit{Identify}, but with the special focus on the time series behind a location factors. In contract, the operations \textit{Associate}, \textit{Correlate}, \textit{Cluster}, or \textit{Distinguish} are not required in this application, whereas \textit{Distribution} is not directly necessary, but automatically given to some extend if the user start the site selection process top -- bottom (\textbf{Task 2}). Summarizing, the most important cognitive operations for site selection are \textit{Identify} and \textit{Compare}.

Bertin \cite{bertin1983semiology} also stated three different \textit{level of reading}, which are elementary (\textit{single data element}), intermediate (\textit{group of elements}), and overall (\textit{all elements together}). Koussoulakou and Kraak \cite{koussoulakou1992spatia} demonstrated that the levels of reading can be independently applied on the spatial and temporal dimensions. In this paper, we prefer to use the term \textit{search level} instead of \textit{level of reading} as suggested by Andrienko et al. \cite{andrienko2003exploratory} due to our focus on exploration of geospatial-temporal data. Furthermore, we unit the intermediate and overall levels into a single category named \textit{general} as suggest by Andrienko et al.

\begin{figure}[htbp]
	\centering
	\includegraphics[width=\columnwidth]{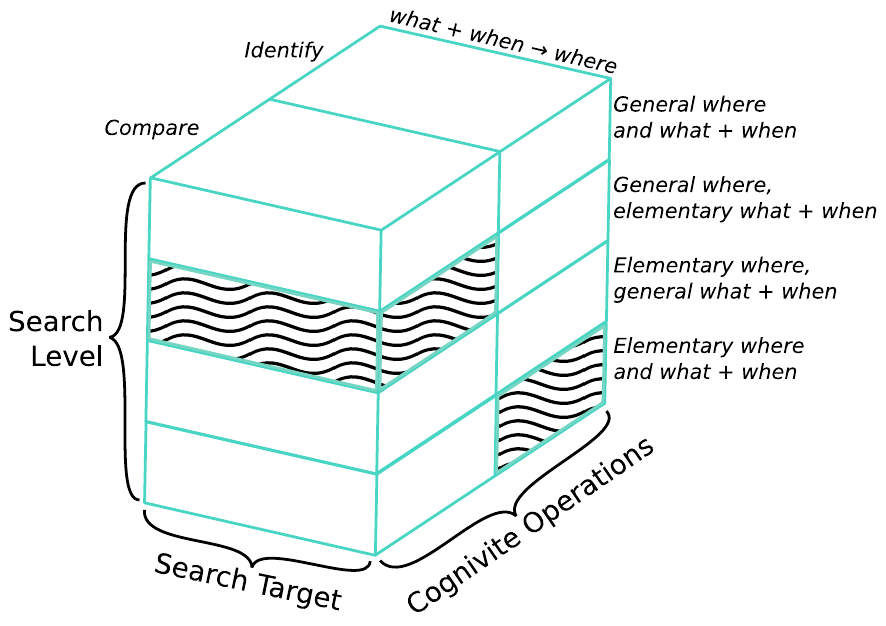}
	\caption{Operational Task Typology Adopted for Site Selection. Shaded Blocks Represents Operation Task of Interest.}
	\label{fig:task_topology}
\end{figure}

Andrienko et al. \cite{andrienko2003exploratory} combined the concepts of \textit{search level} and \textit{search target} with the cognitive operations to propose an operational task typology used to review visualization techniques for geospatial-temporal data. This actually ends up in a five-dimensional classification space for all combinations of search levels, search targets, and cognitive operations. However, in accord with our focus on site selection, this space can be reduced. We focus on the \textit{cognitive operations} Compare and Identify, the \textit{search target} \textit{$when + what \rightarrow where$}, and the \textit{search level} Elementary and General. \autoref{fig:task_topology} shows the resulting operational classification typology schematically.

In this schema, we divided the \textit{search target} into its two components i) \textit{where} \& II) \textit{what + when} and crossbred this division with the elements of \textit{search level}. The resulting four categories are described as follow:

\begin{itemize}
	\item Elementary \textit{where} and elementary \textit{when + what}: describe a site by a given location factor at a given time.
	\item Elementary \textit{where} and general \textit{when + what}: describe site by the change of its location factors over time.
	\item General \textit{where} and elementary \textit{when + what}: describe sites for a given location factor at a given time.
	\item General \textit{where} and general \textit{when + what}: describe sites by the evolution of their location factors over time.
\end{itemize}

In site selection, however, the user perform the selection process top -- down (i.e., from state to region) and thus, starting with assessing all sites (\textit{general}) before narrowing the search window till finally only some sites (\textit{elementary}) are left. Thus, we are interest into the two operational tasks (shaded blocks in \autoref{fig:task_topology}):

\begin{itemize}
	\item General \textit{where} and elementary \textit{when + what}, corresponds to the cognitive operation \textit{Compare}.
	\item Elementary \textit{where} and elementary \textit{when + what}, corresponds to the cognitive operation \textit{Identify}.
\end{itemize}

\subsection{Exploratory Techniques and Operator Primitives}
\label{subsec:Design_exploratory_techniques}

In contrast to paper-based data illustrations, computer-aided visualization tools combine the data representation with user interaction. This enables interactive exploration and dynamic data presentation from which site selection can benefit. This section highlights the techniques, we identified to be valuable and supportive for the given application scenario. More comprehensive information about exploration techniques and work operators can be found in the work of Andrienko et al. \cite{andrienko2003exploratory} and Roth \cite{roth2013empirically}.

We identified two categories of computer-enabled techniques for representation and exploration in the classification of Andrienko et al. \cite{andrienko2003exploratory}. From the perspective of their applicability for site selection, we believe that these techniques are best suited: 

\begin{enumerate}
	\item General techniques, such as querying, map interaction or manipulating views.
	\item Techniques designed for timestamped location factors, such as or time series graph.
\end{enumerate}

The first category contains methods for answering users' questions concerning the data, how are sites described by their location factors in our application case (lookup). This techniques corresponds to the cognitive operation \textit{Identify}. A common tool for enabling access to the geospatial properties of sites are maps. Users can select sites on the map by positioning the mouse cursor over it in order to get insights into the selected site (direct lookup). This additional information can be display in a separate view (focusing). These views, however, must be connected in a way that the information display in each view is providing a coherent picture (linking). Having multiple views entail arbitrarily, but user-defined, layout of these views (arrangement). Focusing, linking, and arranging views were originally suggested by Buja et al. \cite{buja1996interactive}.

Andrienko et al. \cite{andrienko2003exploratory} summarized techniques for showing "temporal information of numeric attribute values at selected locations on a time series-graph". This graph normally have the time on the X-axis and here, the values of the location factor on the Y-axis. Linking the time series-graph to the map, so that the location factors of selected sites are display in the graph, creates an easy to use and simultaneously powerful analysis tool for our geospatial-temporal data. Pointing on a site on the map highlights its values (location factors) in the graph and reversed, selected lines in the graph highlights the corresponding sites. Selecting multiple sites on the map, which is linked to the time series-graph, enables the cognitive operation \textit{Compare}.


\section{Related Work: Visualizing Geospatial-temporal Data}
\label{sec:Related_Work}

To the best knowledge of the authors, no visualization was proposed, which was made for site selection or is suitable for this application scenario. Many approaches for visualizing spatial-temporal data have been published over the year. Sun et al. \cite{sun2016embedding} divided the existing work into two categories: integrated views (the time is combined with the geospatial aspects in the same view) and linked views (the spatial and temporal dimension are displayed in separated, but linked views). 

\textit{Integrating} spatial and temporal information of the data into the same view can be archived by using the well-known space-time cube by Kapler and Wright \cite{kapler2005geotime}. The space is displayed on a map, where the time represents the third dimension. These cubes, however, suffer from the occlusion problem in 3D space. Many different visual approaches were proposed to overcome this issue, such as 2D/3D hybrid representations \cite{tominski2012stacking} having its drawbacks with visualizing data on multiple roads simultaneously, time encoded in the map \cite{liu2011visual} only working for cyclic events, or directly embedding time on top of the spatial data in 2D \cite{aigner2011visualization} leading to occlusion and visual clutter. Another way of integrating space and time are methods for abstraction and aggregation. Researchers evaluated methods that used PCA for transforming the data into abstract space \cite{crnovrsanin2009proximity}, aggregation of movement data for dimensionality reduction \cite{andrienko2008spatio}, or density map where the color represents the time \cite{scheepens2011interactive}.
\newline
However, as described in \autoref{subsec:Design_2D_vs_3D}, we focus on 2D representation of our data and thus, only 2D visualizations are taken into account for site selection. 

\textit{Linked-view} methods are a common approaches for visualizing temporal and spatial data \cite{andrienko2010space}. Ivanov et al. \cite{ivanov2007visualizing} developed a system for efficient monitoring the data of surveillance cameras by combining a timeline, a map, and a camera view. Andrienko et al. \cite{andrienko2011event} used a time series graph in conjunction with a map to visualize several trajectories. Guo et al. \cite{guo2011tripvista} proposed system for analyzing trajectories. His design combines a map for spatial data and a stacked graph along with a scatter plot for temporal aspects in the data. 
\newline
In conclusion, multiple views are a suitable basis for site selection as they show spatial aspects of the data on maps and the temporal information in time series graphs. To reach the best user experiences, the views were often combined and linking (see \autoref{subsec:Design_exploratory_techniques}).  

Additionally, the object of research in the field of visualization is often trajectories, which can be characterized by their changing nature of their spatial properties. Many studies focus on visualizing and analyzing either physical (i.e., walking patterns, routes of buses or taxis) or digital (geo-located tweets from Twitter) activities. More studies about visualizing these trajectories over time can be found in \cite{shaw2009gis, kwan2004geovisualization, kwan2004gis, yu2006spatio}. 
As stated in \autoref{subsec:Design_Data_Classification}, we identified two reasons in \autoref{subsec:Design_Data_Classification} why approaches for visualizing trajectories are not suitable for site selection. First, site selection is all about sites in comparison to point-wise data of trajectories. Second, sites are not moving, i.e., not changing its spatial properties over time. The objectives behind trajectories and site selection differ in principle and thus, visual representation designed for trajectories are not applicable for site selection.

Finally, we determined in \autoref{subsec:Design_Exploration_vs_Analysis} that no dedicated focus on exploratory analysis is necessary in the application field site selection. However, the field of interactively exploration of very large datasets in real time gained special attention. We pointed interested readers to data cubes. As proposed by Stolte et al.'s \cite{stolte2002polaris}, Polaris data cube system was a breakthrough by combining OLAP operations with interactive visualization. Meanwhile, many researchers enhanced the original approach and proposed sophisticated data structures, such as Nancubes \cite{lins2013nanocubes} or Gaussian cubes \cite{wang2017gaussian}, which enable interactive modeling capabilities.

\section{Visual Encoding of Site Selection in QuViS}
\label{sec:Visual_Encoding}

Based on our findings, we have to visually encode geo-referenced location factors for sites. We identified linked views in the literature as most promising baseline for 2D visual encoding of our data. Therefore, we use a map to encode the sites and linked time series graphs to display the location factors to the user. 
\newline
The work flow is defined as follow. First, the user can select location factors from a list of existing factors (known information \textit{what}). As a result, the map will be transformed into a choropleth map coloring each site according to one of the selected location factors. The analyst can explore sites through space and the administrative hierarchy, supporting the cognitive operations \textit{Identify} for the \textit{general} as well as the \textit{elementary} search target top -- bottom. In the next step, a set of sites can be selected the user likes to explore in details. Therefore, the time series graph will show the selected location factors with their full history from which the user can select a particular time point (known information \textit{when}). Details to the preselected sites (unknown information \textit{where}) are shown in a side view attached to the map, where insights of the sites are presented to the user. This facilitate the cognitive operations \textit{Compare}.


\subsection{Interactive Map: Navigation through Space and Administration}
\label{subsec:Visual_Encoding_Map}

\begin{figure}[H]
	\centering
	\subfloat[Lower Bavaria (county level).]{\fbox{\includegraphics[width=\columnwidth]{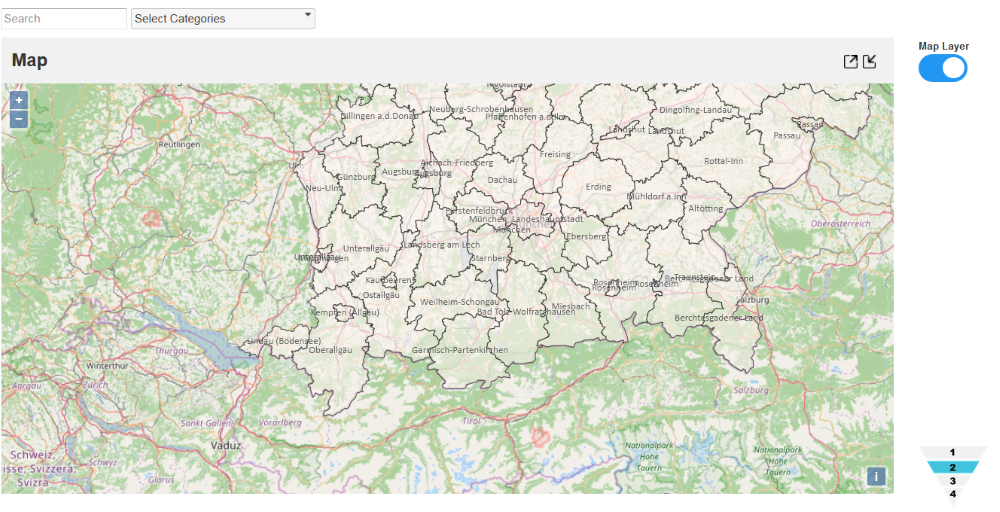}\label{fig:Browse_through_space_a}}}
	\hspace{0cm}
	\subfloat[Garmisch-Partenkirchen (district level).]{\fbox{\includegraphics[width=\columnwidth]{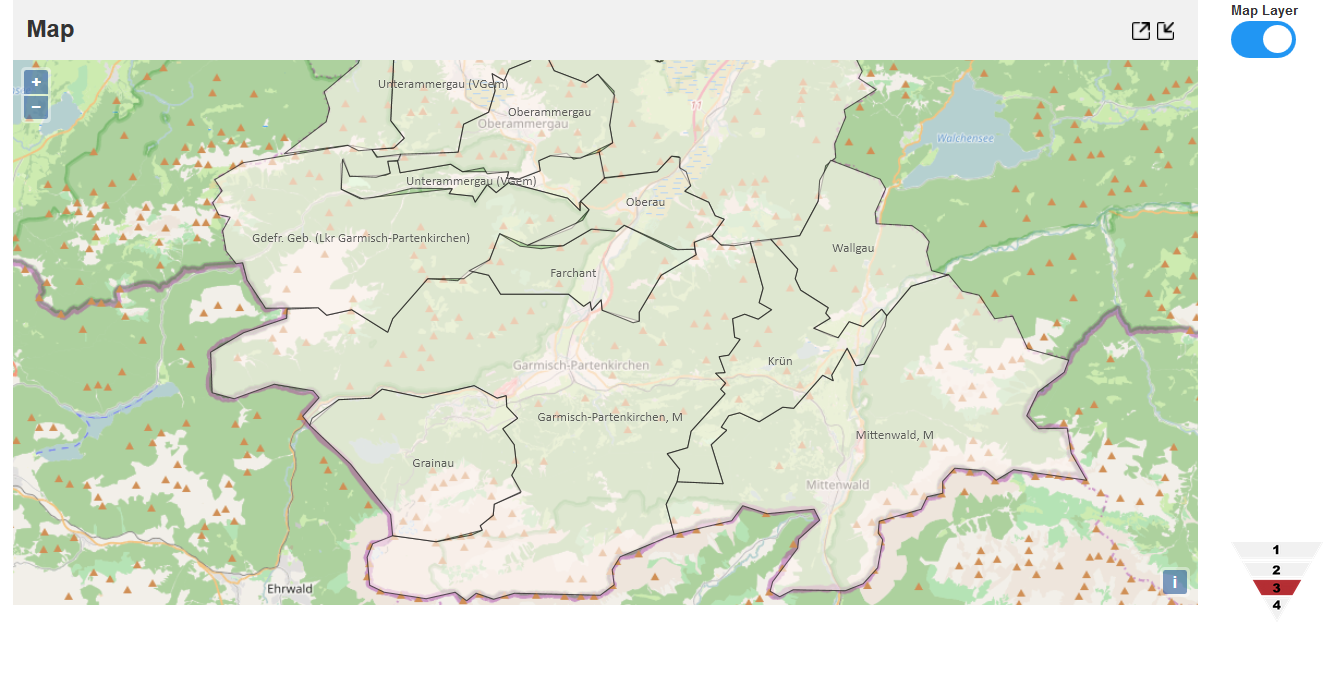}\label{fig:Browse_through_space_b}}}
	\caption{Navigating through Space \& the Administrative Hierarchy.}
	\label{fig:Browse_through_space}
\end{figure}

As the objects of analysis in site selection are sites, the map as a well-known cartographic representation method was chosen to visualize the geospatial dimension. As the sites and the underlying administrative hierarchy refer to a single time moment and do not involve temporal variation, this dimension can be represented and operated as ordinary time-irrelevant spatial data. The map enables users to browse through the geographic dimension. They can easily perceive the spatial properties of sites and in addition, assess their implicit and explicit attributes in a visual fashion. These capabilities address \textit{Task 1} (\autoref{subsec:Background_Conclusion}). \autoref{fig:Browse_through_space} shows an example how users can navigate through space and the administrative hierarchy.

\begin{figure}[!h]
	\centering
	\fbox{\includegraphics[width=\columnwidth]{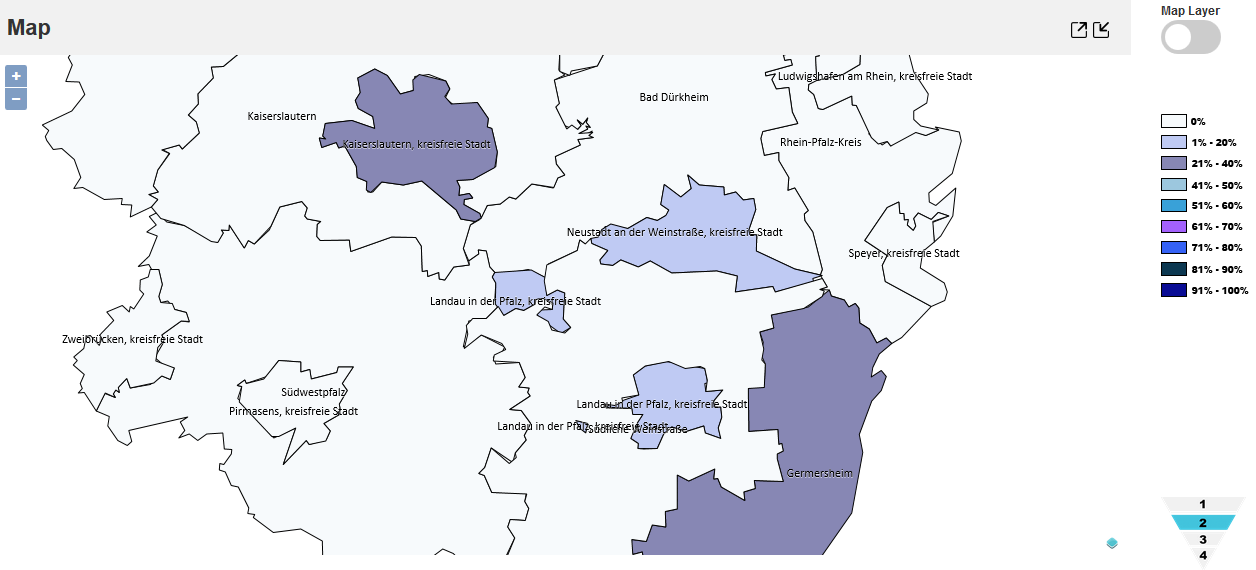}}
	\caption{Visualizing the Inhabitants of Sites with Choropleth Maps.}
	\label{fig:Choropleth_Map}
\end{figure}

Furthermore, users can navigate through the administrative hierarchy, addressing \textit{Task 2}. In this hierarchy, each site has one parent site \footnote{Currently, Germany is the root node and thus, do not contain a parent.} and a specific set of child site\footnote{The municipality level is currently the most granular level in the hierarchy.}. Hovering over a site highlights this particular site with a blue border. If he double clicks on this site, he drill down the hierarchy and the sites' child sites are presented. By right clicking on the map, a roll up is initiated. The parent node with all of its neighbors within this level in the hierarchy is shown. For instance, if a user right clicks in the county level, all 16 states of Germany are presented. 
\newline
With this interactive map, decision-makers can identify sites, determine its spatial characteristics and is able to compare them with other sites on the elementary as well as the general \textit{search level}. Additionally, we utilize choropleth maps in order to directly display location factors on the map (addressing \textit{Task 3}). Here, sites as the spatial units are entirely colored according to a user preselected location factor, such as inhabitants or unemployment rates. Choropleth maps provide an easy way to visually analyze how a location factors varies across a geographic region. \autoref{fig:Choropleth_Map} depicts the inhabitants of exemplary selected countries of the state \textit{Rhineland-Palatinate}.

Concluding, users can visually analyze location factors with choropleth maps as well as navigate through space and the administrative hierarchy. Individual sites (search level elementary) as well as the overall characteristics of sites (search level general) can be assessed.

\subsection{Exploratory Graphs: Comparing Location Factors over Time}
\label{subsec:Visual_Encoding_Time_Series}

\textit{Task 3} implies in conjunction with \textit{Task 5} that an analyst have to be able to visualize location factors referring to a particular user-selected time moment. Although he judges sites for their current characteristics, analysts must also take the history into account in order to rate the sites' future development. This eventually enables decision-makers to determine sites where the sites' corresponding location factors fulfill the requirements of the company -- for now and presumably in the future.
\newline
Therefore, we support elaborated querying facilities. The location factor of interest as well as the user selected time point (known information) can be set as query constraints, and characteristics of sites (unknown information) is the query target. This way, users can \textit{identify} sites on the \textit{elementary} level (with respect to sites) and get descriptions of their geospatial properties together with their location factors displayed on the map. The same procedure can be also applied for \textit{comparing} multiple sites on the \textit{elementary} level where users need to see the differences in the sites' location factors at the same time or the changes of the location factors at different time points.

\begin{figure}[!h]
	\centering
	\fbox{\includegraphics[width=\columnwidth]{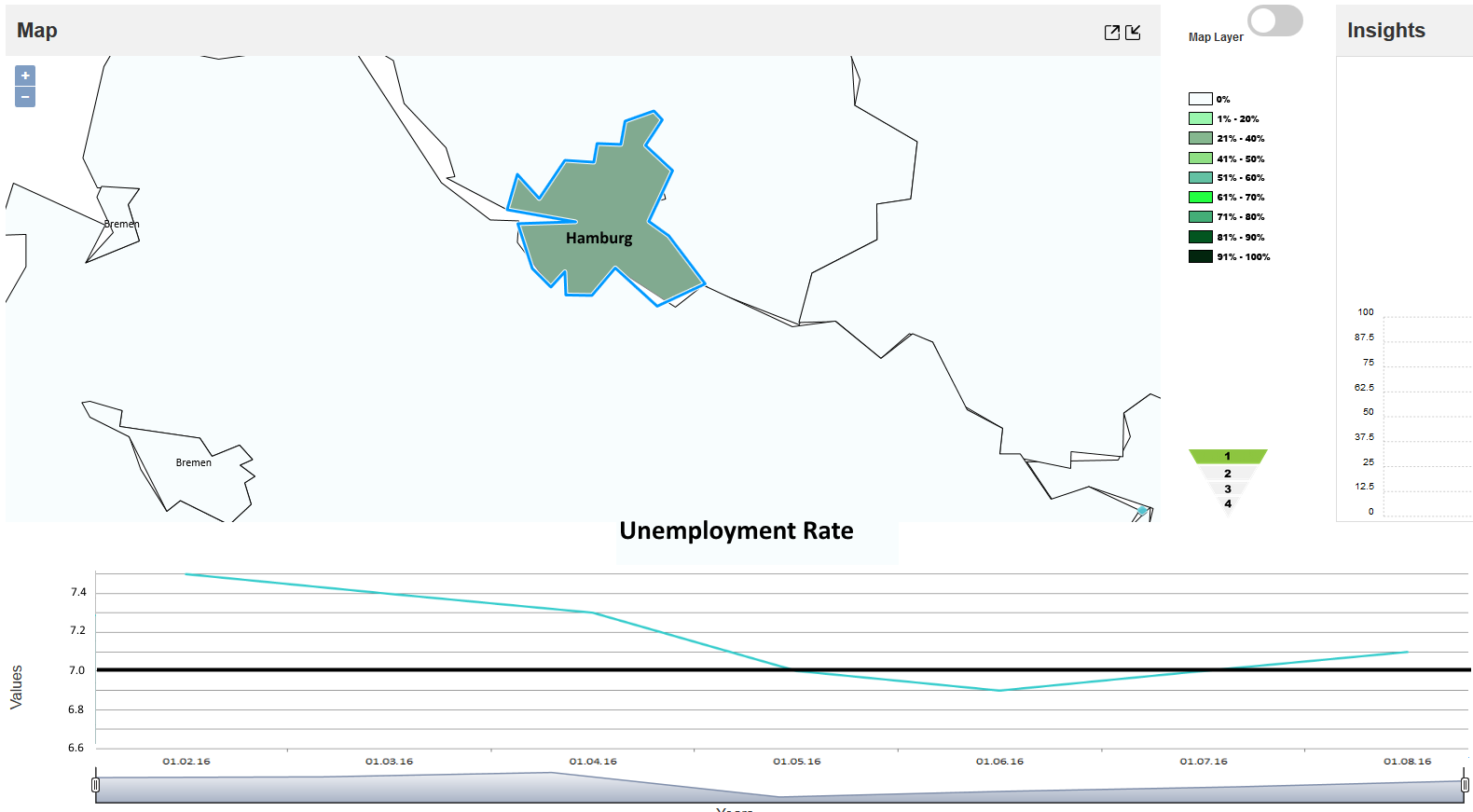}}
	\caption{Interactive Time Series Graph for Identifying Individual Sites based on their Location Factors.}
	\label{fig:TS_Identifying}
\end{figure}

As suggested by Andrienko et al. \cite{andrienko2003exploratory}, we applied interactive time-series graph to facilitate visual identification and comparison tasks as the location factors are numeric by nature (see \autoref{sec:Related_Work}).
\newline
In order to answer questions like "When was the unemployment rate in Hamburg below the level of 7 percent?", analysts can use the 7 percent as an reference value and add a straight horizontal line to the graph by hover over the axe at this value. The user can select Hamburg by clicking on it in the map to see if there unemployment rate have been below 7 percent (see \autoref{fig:TS_Identifying}). It can be easily seen that the unemployment rate in Hamburg was below 7 percent in June 2016. 

\begin{figure}[!h]
	\centering
	\fbox{\includegraphics[width=\columnwidth]{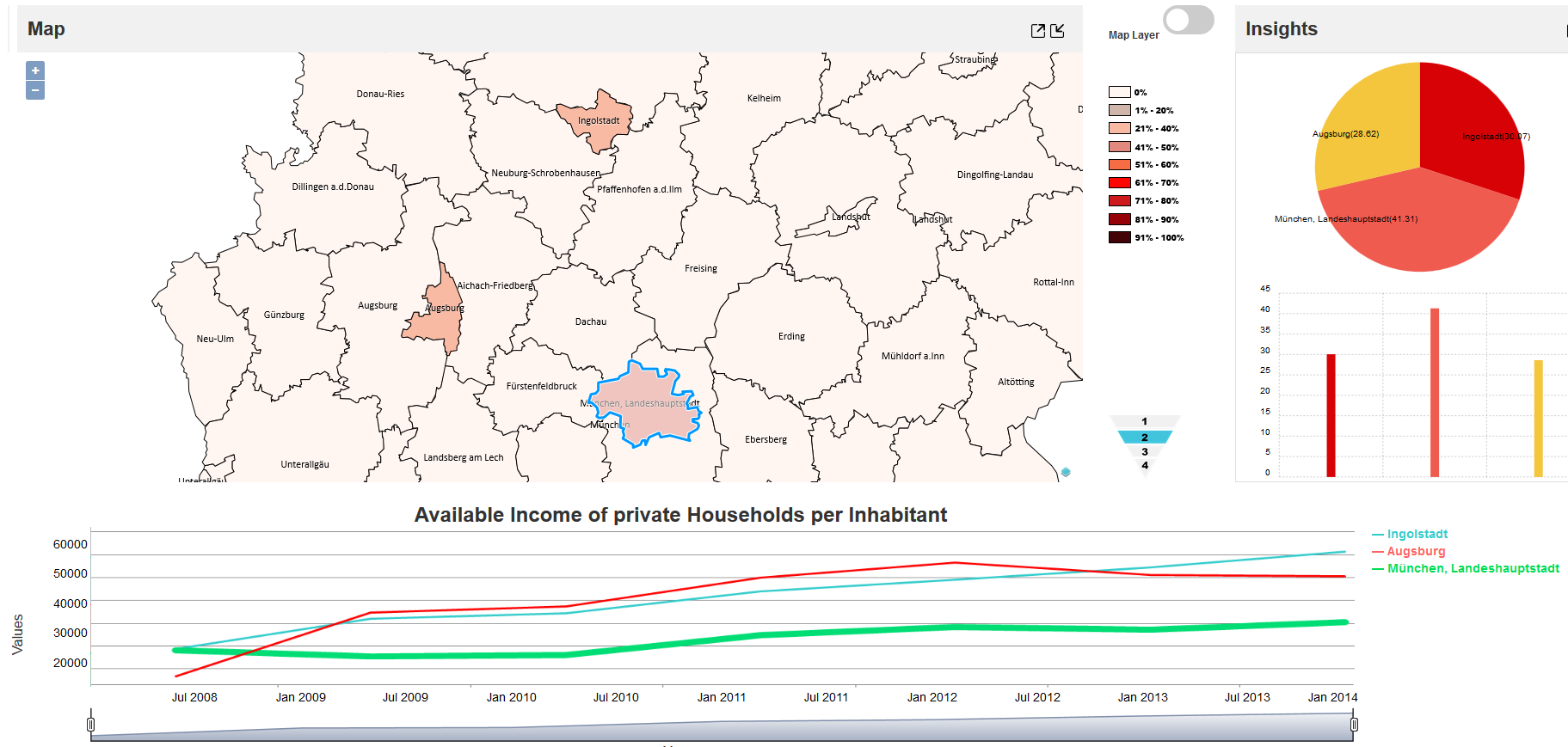}}
	\caption{Interactive Time Series Graph for Identifying Individual Sites based on their Location Factors.}
	\label{fig:TS_Comparing}
\end{figure}

The time series graph is well suitable to support comparisons of sites with a particular reference sites to answer questions like "Where and when was the purchasing power higher than in Munich?". For these comparisons, a particular site respective its corresponding values are highlighted in the time series graph. Again, the user has to select Munich from the map and compare other sites with the highlighted line in the graph. \autoref{fig:TS_Comparing} present the time series graph for this example with the corresponding map.

However, the applied time series graph has two limitations. First, the graph only represents values for one location factors. Although multiple location factors can be theoretically stacked in order to display more factors, this approach does not only covers too much window space, but also tends to be confusing. Second, the graph is only suitable to present the characteristics of individual sites separately. It cannot depict the overall development of regions as a whole (search level general).

\subsection{Details on Demand: Viewing Insights on Sites}
\label{subsec:Visual_Encoding_Insights}

\begin{figure}[!h]
	\centering
	\fbox{\includegraphics[width=0.4\columnwidth]{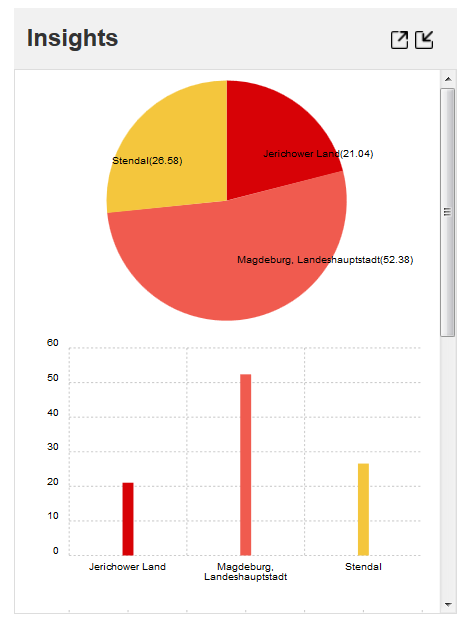}}
	\caption{The Side View \textit{Insights} .}
	\label{fig:Insights_Charts}
\end{figure}

To facilitate the comparison of sites, we have integrated two additional views to support decision-makers in the \textit{operational tasks} of comparing sites in general for a given time point and location factor. Therefore, \textit{QuViS} supports the view \textit{insights} for visualizing data in charts and diagrams as well as the view \textit{data table} for showing the raw data.
\newline
We designed the side view \textit{insight} for displaying comparisons of the selected sites for the and thus, helping users to see the differences of sites for the given location factors visually. \autoref{fig:Insights_Charts} shows an example for three selected countries\footnote{Magdeburg is a urban district, whereas Jerichower Land and stendal are rural districts.} within the state \textit{Saxony-Anhalt}. First, a pie chart is generated which is divided into slices to illustrate the numerical proportion of the location factors for each selected sites. Additionally, a bar chart is shown where each bar represents a location factors with the bar's length being proportional to the value for the corresponding site. Although, we focused on facilitating visual comparison of sites' location factors, more insights can be integrated and shown to users in order to support them in decision making. Statistical reports can be computed utilizing the fact that each selected site has usually a set of child sites for which the average, minimum, maximum, and deviation are derivable. More sophisticated extensions will use data mining methods to generate new insights of the data to not only show the data, but also present unknown information to users. We point interested readers to our previous work \cite{baumbach2017QuIS} how to identify hidden influencing location factors.

\begin{figure}[!h]
	\centering
	\fbox{\includegraphics[width=\columnwidth]{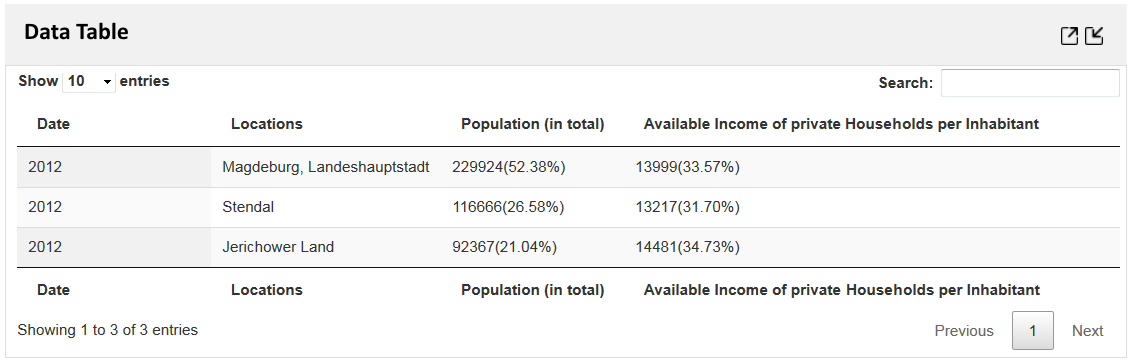}}
	\caption{The Side View \textit{Data Table} lists the Raw Values of the Selected Location Factors.}
	\label{fig:Insights_Tables}
\end{figure}

Furthermore, we present the raw values of the chosen location factors to the user for the selected location factors and time point. This way, we enable decision makers in assessing the corresponding values by themselves in order to form their own opinion or expert his information for the future use. \autoref{fig:Insights_Tables} shows the raw values for already presented example of \autoref{fig:Insights_Charts}.

\section{Use-Case Scenario: German Supermarkets}
\label{sec:case_study}

We demonstrate the capabilities of our design on a case study of supermarkets in Germany. Experts have described the relevant location factors which form the basis for supermarket chains performing site selection \cite{greiner1997standortbewertung}. According to their expansion strategies, \emph{Edeka} (operating both Edeka Supermarkets and E-Center), \emph{Lidl}, and \emph{NP} supermarket chains assert to apply the same criteria\footnote{Source: \href{http://www.lidl-immobilien.de/cps/rde/xchg/SID-6DF6A32A-8E7E665B/lidl_ji/hs.xsl/5186.htm}{www.lidl.de} and \href{https://www.edeka-verbund.de/Unternehmen/de/edeka_minden_hannover/immobilien_minden_hannover/expansion_minden_hannover/edeka_expansion_minden_hannover.jsp?}{www.edeka-verbund.de}.}. The main criterion is the location factor \textit{inhabitants}. The additional, external factors, such as sales floor, plot area, availability of parking places, rents, or sales floor of available real estate, are beyond the scope of this paper (see \autoref{sec:Background}). The results of our previous study for \textit{QuIS} showed, however, that the supermarkets chains also take the purchasing power of people into consideration \cite{baumbach2017QuIS}.
\newline
The regional focus, defined by the sphere of influence of NP, covers $1,704$ municipalities respective cities with a total amount of $1,556$ existing supermarkets. Within this area, we were looking for sites which fulfill the given requirements of the supermarket chains (listed in \autoref{tab:SMLocationReq}). Furthermore, the \textit{number of inhabitants} as well as the \textit{purchasing power} are additional requirements which were to maximize for a potential sites.

\begin{table}[!h]
	\def\arraystretch{1.25}
	\caption{Location Requirements for Supermarkets}
	\label{tab:SMLocationReq}
	\centering
	\resizebox{1\columnwidth}{!}{
		\begin{tabular}{l|c|c|c|c}
			& Edeka          & E-Center        & Niedrig Preis (NP) & Lidl        \\ 
			\hline
			Core Population      & 5.000          & 10.000          & 2.500              & 5.000       \\
			\hline
			Area Population & 8.000          & 25.000          & 5.000              & 10.000      \\ 
		\end{tabular}
	}
\end{table}

\subsection{Data Model and Dataset}
\label{subsec:case_study_data_model}

\begin{figure}[h]
	\centering
	\includegraphics[width=0.75\columnwidth]{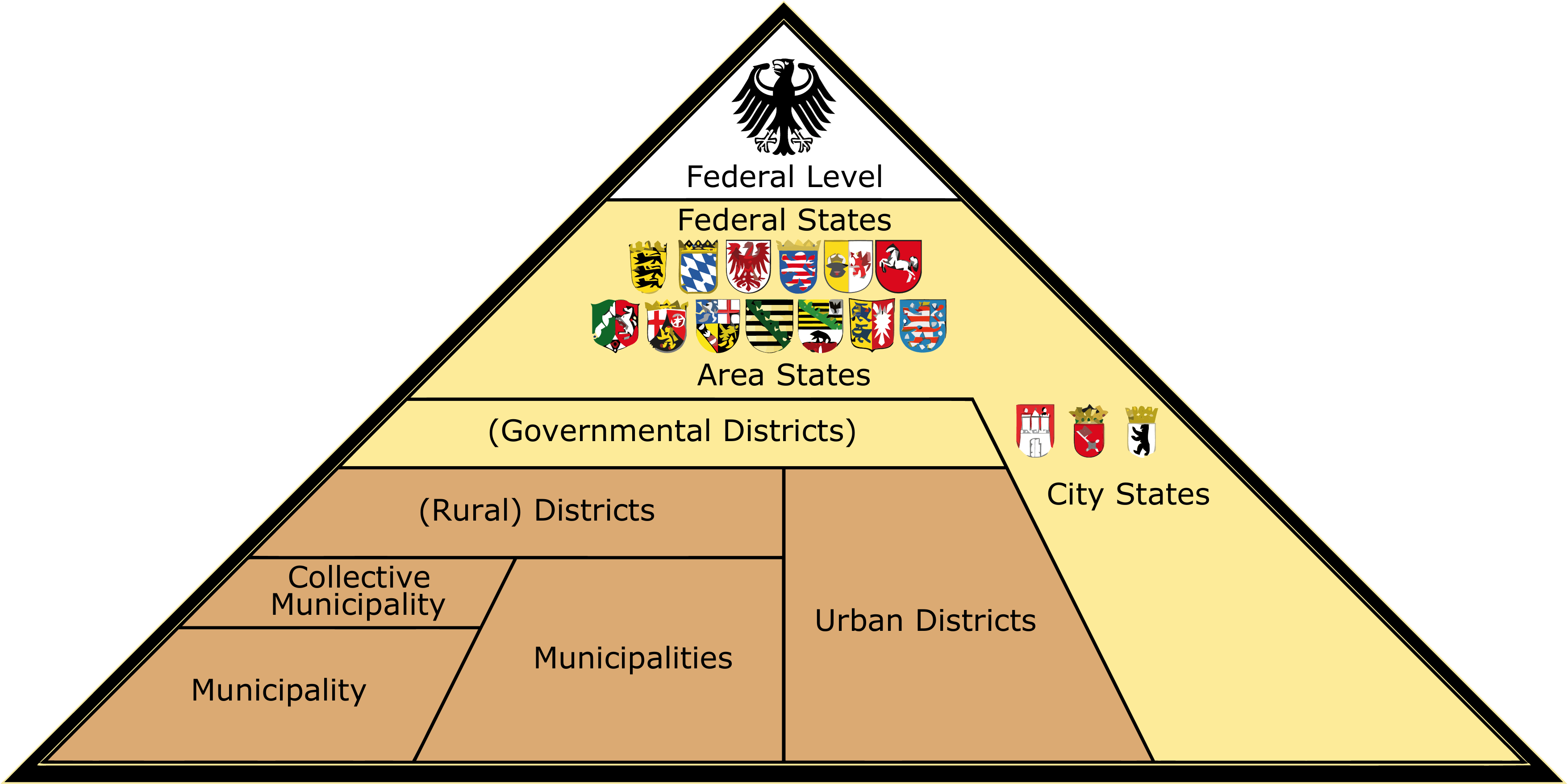}
	\caption[Administrative divisions of Germany.]{Administration Hierarchy Germany, By David Liuzzo \href{http://creativecommons.org/licenses/by-sa/2.0}{CC BY-SA 2.0}.}
	\label{fig:hierarchy}
\end{figure}

In order to perform exploratory visual site selection, a data model capturing sites as well as location factors is necessary. The data model incorporates geospatial-temporal data from various sources and structures it in a geographical, temporal, and hierarchical way. Interested readers find a comprehensive description of the utilized data model in our previous work \cite{baumbach2017QuIS}.
\newline
The data model organizes all sites (here, of Germany) in a hierarchy according to a nation's federalization. \autoref{fig:hierarchy} shows the structure of the data model where each state, county, district, and municipality is represented as a single site. In addition to these sites, the data model contains the location factors, which are technically modeled as geo-referenced time series as described in \autoref{subsec:Design_Data_Classification}. 

For this paper, we have collected data for all $16$ states, $402$ counties, $4,520$ districts, and $11,162$ municipalities as well as information about the hierarchy of Germany. This data is provided by German Federal Statistical Office. Additionally, the data model contains $450$ location factors for the last $15$ to $20$ years (depending on the specific location factor). These location factors contains, among others, net income, purchasing power, various information about inhabitants \& population, employees \& unemployed persons, education, land costs, households in number \& size, and companies itemized in size, number, profit, industry, \& employed people. 

\subsection{Visually Data Analysis for Site Selection}
\label{subsec:case_study_visually_Analysis}

\begin{figure}[!h]
	\centering
	\fbox{\includegraphics[width=\columnwidth]{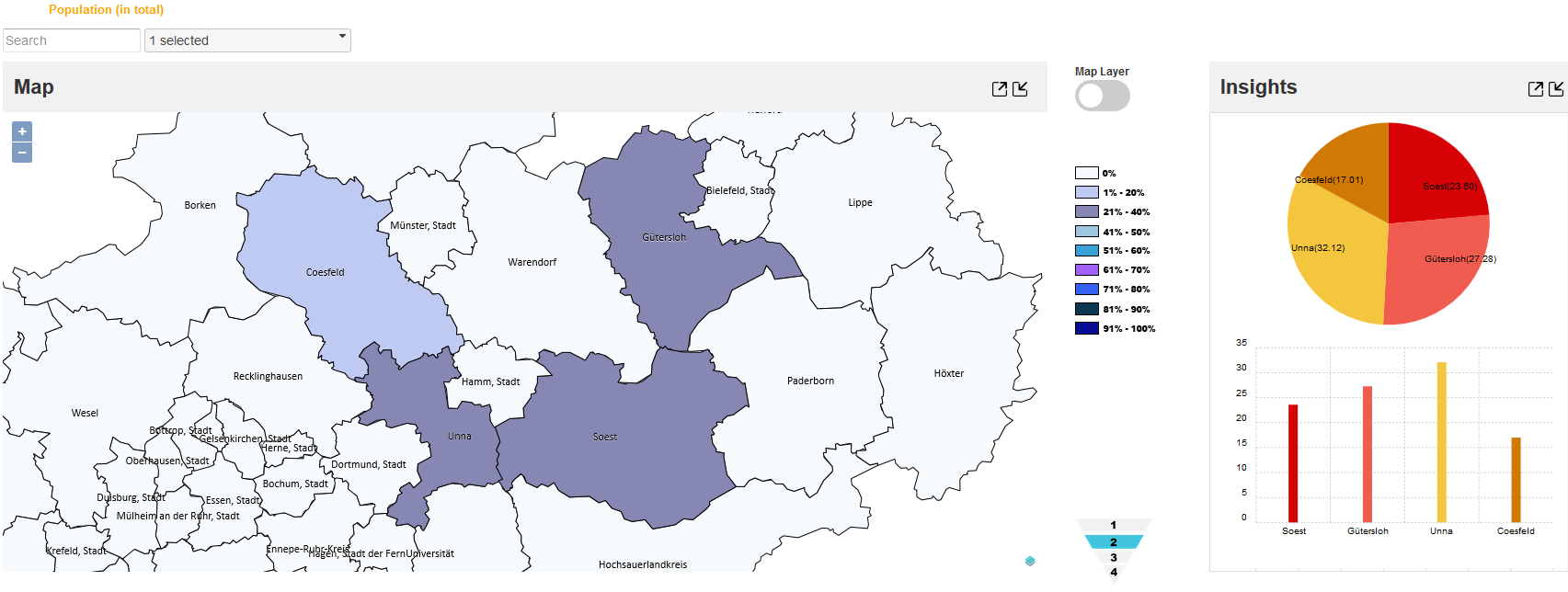}}
	\caption{Population (in total) for Selected Countries.}
	\label{fig:NRW_Population}
\end{figure}

In our study for the site recommender system \textit{QuIS}, one of the suggested sites for a new supermarket was the district \textit{Herzebrock-Clarholz} in the county \textit{G\"utersloh}. Although, this site was automatically recommended by\textit{QuIS}, the decision-makers could have also selected this site manually. Manually selecting sites implies that the user had to be able to browse through all existing sites to explore their location factors interactively. Automated recommendation of sites, however, also raises the demand of exploratory data analysis to provide explanations about the automatically suggested sites. Otherwise, decision-makers in companies will not trust the recommendations if the system acts as a black box \cite{baumbach2017QuIS}. Therefore, we like to demonstrate the exploratory visual analysis capabilities of \textup{QuViS} by analyzing why \textit{G\"utersloh} is a suitable site for a new supermarket.
\newline
As stated above, we are interested in the location factors \textit{inhabitants}, \textit{purchasing power} and \textit{existing supermarkets} (as part of a competitor analysis). Site selection is done top -- bottom, so we are looking at the population for each county within the state \textit{Nordrhein-Westfalen} first. We have selected four exemplary countries, namely \textit{Coesfeld}, \textit{G\"utersloh},  \textit{Unna}, and \textit{Soest}. \autoref{fig:NRW_Population} shows visualized population for these countries.

User easily perceive that the country \textit{Unna} has the highest population count (416,679 inhabitants), followed by \textit{G\"utersloh} (353,944), \textit{Soest} (306,131), and \textit{Coesfeld} (220,662). Based on the location factor \textit{population}, the country \textit{Unna} is of interest for supermarkets. 

\begin{figure}[!h]
	\centering
	\fbox{\includegraphics[width=\columnwidth]{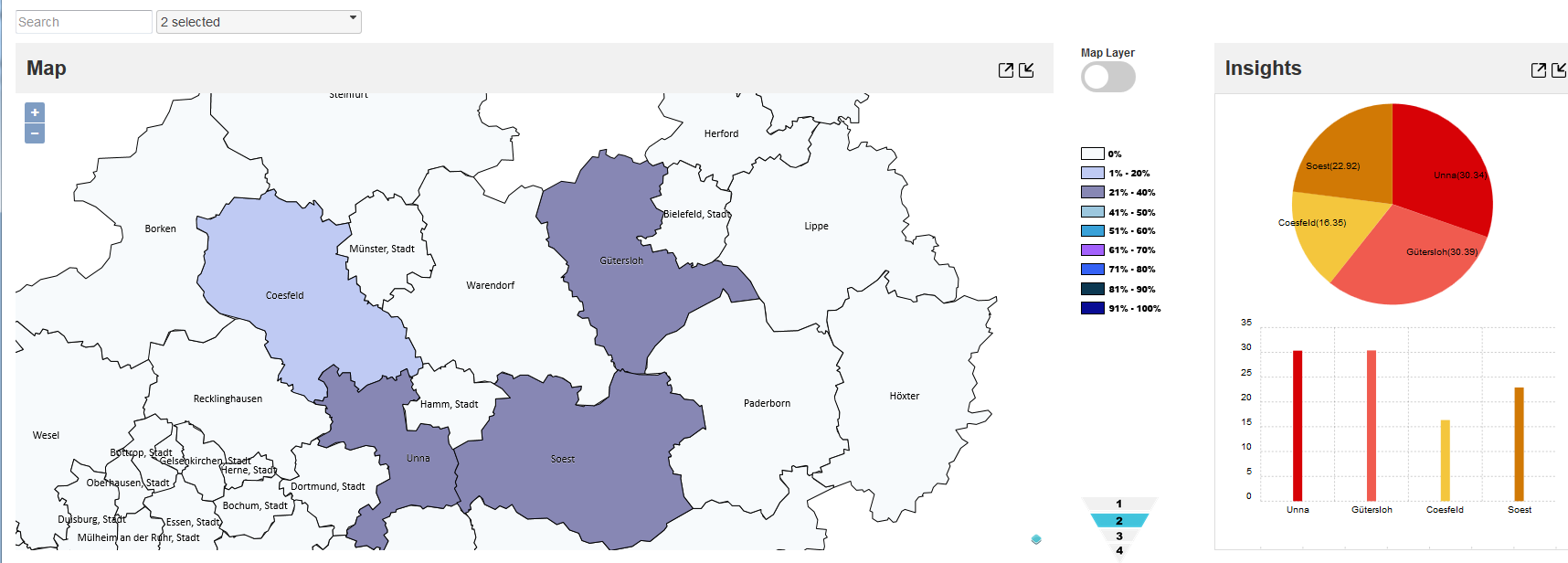}}
	\caption{Available Income per private Household for Selected Countries.}
	\label{fig:NRW_Income}
\end{figure}

Next, we analyze the available income per private household. \autoref{fig:NRW_Income} shows the visualization of this location factor for the selected countries. Here, we see that \textit{G\"utersloh} has the highest income per private household (\EUR{18,102}) and the highest income of private households (\EUR{6,229,891}) as well. The country \textit{Unna} has the same amount of income of private households (\EUR{6,219,899}), but as it has more inhabitants, the income per private household is lower with (\EUR{14,451}). So we select \textit{G\"utersloh} as it has both, many inhabitants and a high income per inhabitant.

Within the country \textit{G\"utersloh}, we are also looking for already existing supermarkets in the districts to avoid competition. From \autoref{fig:NRW_Income}, it can be seen that the district \textit{Herzebrock-Clarholz} is a suitable site for a supermarket. \textit{Herzebrock-Clarholz} has $6.497$ private households with $15,969$ inhabitants in total as well as a high average, available income per household of \EUR{53.310}. Most important, however, no competing supermarket exists in this districts, so that \textit{Herzebrock-Clarholz} represents a suitable site for a new supermarket.

\begin{figure}[!h]
	\centering
	\fbox{\includegraphics[width=\columnwidth]{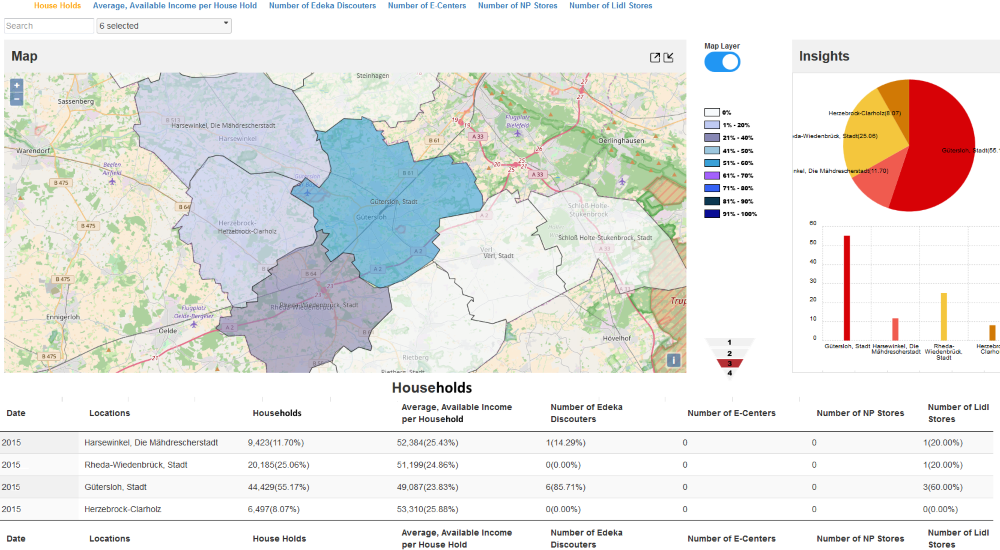}}
	\caption{Number of Households for selected Districts in the Country \textit{G\"utersloh}.}
	\label{fig:Gütersloh_Supermarkets}
\end{figure}

\section{Conclusion and Future Work}
\label{sec:conclusion}

In this paper, we explored the use exploratory visual data analysis for site selection and designed a system for interactive exploration of geo-referenced location factors over time. We demonstrated in the use case scenario of selecting a suitable site for a supermarket in Germany the visualization and exploration capabilities of our presented visual representation. Existing approaches for visualizing spatial-temporal data focus on point-wise data, such as tweets, or trajectories of humans where the spatial objects are constantly moving over time. 
\newline
As Andrienko et al. described in their work \cite{andrienko2003exploratory}, spatial-temporal data can change over time in different ways according to the occurring changes. In our paper, we focused on "changes of thematic properties through values of attributes", i.e., location factors in this application scenario, only. There are, however, also existential changes (where sites can appear respective disappear) as well as changes of spatial properties (with regard to the sites' location, shape, and size). One example is consolidations, which happing all over the world. In the state Bavaria, Germany, the number of municipalities dropped from roughly 7,000 to roughly 2,000 between 1972 and 1978. In 1990, the former East Germany dissolved and acceded to the Federal Republic. Political initiated consolidations always follow spatial changes, as the merged entity obtains a united shape, new centroid coordinates, and an updated sizes. Modeling these changes in the administrative hierarchy is not only an end in itself. All location factors are structured as time series and their values are linked to specific sites within the administrative hierarchy. Currently, no time series values for sites can be imported into the data model as these sites do not currently exist. To solve this problem, merging, separating, and dispersing also have to be modeled in time series.
\newline
Furthermore, social media can be also integrated into the proposed data model as another source of information, which supports decision-makers in site selection. Social networking sites like Facebook or microblogging services, such as Twitter, gained a great popularity worldwide. This lead, in conjunction with ubiquitous and widespread mobile devices, to a constantly growing volume of geo-referenced and timestamped data. 
Collecting and analyzing social media data in this way enables users to directly analyze these user demands within the site selection process. Similar to the proposed approach of Jaffe et al. \cite{jaffe2006generating} for summarizing large-scale collections of geo-referenced photo collections, well-known tag clouds can be adopted to integrate user demands into the proposed visualization system \cite{hassan2006improving}. 
\newline
The same way as social media can be integrated into the visualization system, supplier, customer, and competitor analysis can be incorporated to enhance the selection process. Yellow pages contain detailed and overall information about companies with their stores, factories and logistic centers. Having this information imported into the data model as another data source, the distribution of companies within the catchment area of a particular site can be aggregated and analyzed. As a consequence, the attractiveness of this particular site for a given company can be calculated when the number of existing competitors, potential customers and available supplier is known a priori.






\bibliographystyle{abbrv}
\bibliography{paper}

\begin{thebibliography}{10}

\bibitem{aigner2011visualization}
W.~Aigner, S.~Miksch, H.~Schumann, and C.~Tominski.
\newblock {\em Visualization of time-oriented data}.
\newblock Springer Science \& Business Media, 2011.

\bibitem{andrienko2008spatio}
G.~Andrienko and N.~Andrienko.
\newblock Spatio-temporal aggregation for visual analysis of movements.
\newblock In {\em Visual Analytics Science and Technology, 2008. VAST'08. IEEE
  Symposium on}, pages 51--58. IEEE, 2008.

\bibitem{andrienko2010space}
G.~Andrienko, N.~Andrienko, U.~Demsar, D.~Dransch, J.~Dykes, S.~I. Fabrikant,
  M.~Jern, M.-J. Kraak, H.~Schumann, and C.~Tominski.
\newblock Space, time and visual analytics.
\newblock {\em International Journal of Geographical Information Science},
  24(10):1577--1600, 2010.

\bibitem{andrienko2011event}
G.~Andrienko, N.~Andrienko, and M.~Heurich.
\newblock An event-based conceptual model for context-aware movement analysis.
\newblock {\em International Journal of Geographical Information Science},
  25(9):1347--1370, 2011.

\bibitem{andrienko2006exploratory}
N.~Andrienko and G.~Andrienko.
\newblock {\em Exploratory Analysis of Spatial and Temporal Data: A Systematic
  Approach}.
\newblock Springer-Verlag New York, Inc., Secaucus, NJ, USA, 2005.

\bibitem{andrienko2003exploratory}
N.~Andrienko, G.~Andrienko, and P.~Gatalsky.
\newblock Exploratory spatio-temporal visualization: an analytical review.
\newblock {\em Journal of Visual Languages \& Computing}, 14(6):503--541, 2003.

\bibitem{badri2007dimensions}
M.~A. Badri.
\newblock Dimensions of industrial location factors: review and exploration.
\newblock {\em Journal of Business and Public Affairs}, 1(2):1--26, 2007.

\bibitem{Bankhofer2001}
U.~Bankhofer.
\newblock {\em Industrial Location Management}.
\newblock German university-Verlag, 2001.

\bibitem{baumbach2017QuIS}
S.~Baumbach, F.~Sachs, S.~Ahmed, and A.~Dengel.
\newblock Quis: The question of intelligent site selection.
\newblock {\em Submitted to IEEE Transactions on Knowledge and Data
  Engineering}, 2017.
\newblock http://www.dfki.uni-kl.de/~baumbach/dl/paper.pdf.

\bibitem{bertin1983semiology}
J.~Bertin.
\newblock Semiology of graphics: diagrams, networks, maps.
\newblock 1983.

\bibitem{Bhatnagar2005443}
R.~Bhatnagar and A.~S. Sohal.
\newblock Supply chain competitiveness: measuring the impact of location
  factors, uncertainty and manufacturing practices.
\newblock {\em Technovation}, 25(5):443 -- 456, 2005.

\bibitem{blair1987major}
J.~P. Blair and R.~Premus.
\newblock Major factors in industrial location: A review.
\newblock {\em Economic Development Quarterly}, 1(1):72--85, 1987.

\bibitem{buja1996interactive}
A.~Buja, D.~Cook, and D.~F. Swayne.
\newblock Interactive high-dimensional data visualization.
\newblock {\em Journal of computational and graphical statistics}, 5(1):78--99,
  1996.

\bibitem{crnovrsanin2009proximity}
T.~Crnovrsanin, C.~Muelder, C.~Correa, and K.-L. Ma.
\newblock Proximity-based visualization of movement trace data.
\newblock In {\em Visual Analytics Science and Technology, 2009. VAST 2009.
  IEEE Symposium on}, pages 11--18. IEEE, 2009.

\bibitem{glatte2015location}
T.~Glatte and T.~Haupt.
\newblock Location strategies: methods and their methodological limitations.
\newblock {\em Journal of Engineering, Design and Technology}, 13(3), 2015.

\bibitem{gray1997data}
J.~Gray, S.~Chaudhuri, A.~Bosworth, A.~Layman, D.~Reichart, M.~Venkatrao,
  F.~Pellow, and H.~Pirahesh.
\newblock Data cube: A relational aggregation operator generalizing group-by,
  cross-tab, and sub-totals.
\newblock {\em Data mining and knowledge discovery}, 1(1):29--53, 1997.

\bibitem{greiner1997standortbewertung}
H.~Greiner.
\newblock Standortbewertung im einzelhandel?organisation und durchf{\"u}hrung
  der standortselektion am beispiel der rewe-gruppe.
\newblock In {\em Handelsforschung 1997/98}, pages 233--253. Springer, 1997.

\bibitem{guo2011tripvista}
H.~Guo, Z.~Wang, B.~Yu, H.~Zhao, and X.~Yuan.
\newblock Tripvista: Triple perspective visual trajectory analytics and its
  application on microscopic traffic data at a road intersection.
\newblock In {\em Pacific Visualization Symposium (PacificVis), 2011 IEEE},
  pages 163--170. IEEE, 2011.

\bibitem{hassan2006improving}
Y.~Hassan-Montero and V.~Herrero-Solana.
\newblock Improving tag-clouds as visual information retrieval interfaces.
\newblock In {\em International conference on multidisciplinary information
  sciences and technologies}, pages 25--28, 2006.

\bibitem{hicks2003comparison}
M.~Hicks, C.~O'Malley, S.~Nichols, and B.~Anderson.
\newblock Comparison of 2d and 3d representations for visualising
  telecommunication usage.
\newblock {\em Behaviour \& Information Technology}, 22(3):185--201, 2003.

\bibitem{ivanov2007visualizing}
Y.~Ivanov, C.~Wren, A.~Sorokin, and I.~Kaur.
\newblock Visualizing the history of living spaces.
\newblock {\em IEEE Transactions on Visualization and Computer Graphics},
  13(6):1153--1160, 2007.

\bibitem{jaffe2006generating}
A.~Jaffe, M.~Naaman, T.~Tassa, and M.~Davis.
\newblock Generating summaries and visualization for large collections of
  geo-referenced photographs.
\newblock In {\em Proceedings of the 8th ACM International Workshop on
  Multimedia Information Retrieval}, MIR '06, pages 89--98, New York, NY, USA,
  2006. ACM.

\bibitem{kapler2005geotime}
T.~Kapler and W.~Wright.
\newblock Geotime information visualization.
\newblock {\em Information Visualization}, 4(2):136--146, 2005.

\bibitem{kaya20143d}
E.~Kaya, M.~T. Eren, C.~Doger, and S.~Balcisoy.
\newblock Do 3d visualizations fail? an empirical discussion on 2d and 3d
  representations of the spatio-temporal data, 2014.

\bibitem{Kisilevich2010}
S.~Kisilevich, F.~Mansmann, M.~Nanni, and S.~Rinzivillo.
\newblock {\em Spatio-temporal clustering}, pages 855--874.
\newblock Springer US, Boston, MA, 2010.

\bibitem{kjellin2010evaluating}
A.~Kjellin, L.~W. Pettersson, S.~Seipel, and M.~Lind.
\newblock Evaluating 2d and 3d visualizations of spatiotemporal information.
\newblock {\em ACM Trans. Appl. Percept.}, 7(3):19:1--19:23, June 2008.

\bibitem{koua2006evaluating}
E.~L. Koua, A.~MacEachren, and M.-J. Kraak.
\newblock Evaluating the usability of visualization methods in an exploratory
  geovisualization environment.
\newblock {\em International Journal of Geographical Information Science},
  20(4):425--448, 2006.

\bibitem{koussoulakou1992spatia}
A.~Koussoulakou and M.-J. Kraak.
\newblock Spatia-temporal maps and cartographic communication.
\newblock {\em The cartographic journal}, 29(2):101--108, 1992.

\bibitem{kwan2004gis}
M.-P. Kwan.
\newblock {GIS} {M}ethods in {T}ime-{G}eographic {R}esearch: {G}eocomputation
  and {G}eovisualization of {H}uman {A}ctivity {P}atterns.
\newblock {\em Geografiska Annaler: Series B, Human Geography}, 86(4):267--280,
  2004.

\bibitem{kwan2004geovisualization}
M.-P. Kwan and J.~Lee.
\newblock Geovisualization of {H}uman {A}ctivity {P}atterns using {3D GIS}: a
  time-geographic approach.
\newblock {\em Spatially integrated social science}, 27, 2004.

\bibitem{lehr1885mathematische}
J.~Lehr and W.~Launhardt.
\newblock Mathematische begr{\"u}ndung der volkswirtschaftslehre, 1885.

\bibitem{liebmann1969grundlagen}
H.-P. Liebmann.
\newblock {\em Grundlagen betriebswirtschaftlicher Standortentscheidungen}.
\newblock na, 1969.

\bibitem{lins2013nanocubes}
L.~Lins, J.~T. Klosowski, and C.~Scheidegger.
\newblock Nanocubes for real-time exploration of spatiotemporal datasets.
\newblock {\em IEEE Transactions on Visualization and Computer Graphics},
  19(12):2456--2465, 2013.

\bibitem{liu2011visual}
H.~Liu, Y.~Gao, L.~Lu, S.~Liu, H.~Qu, and L.~M. Ni.
\newblock Visual analysis of route diversity.
\newblock In {\em Visual Analytics Science and Technology (VAST), 2011 IEEE
  Conference on}, pages 171--180. IEEE, 2011.

\bibitem{luder1983unternehmerische}
K.~L{\"u}der and W.~K{\"u}pper.
\newblock {\em Unternehmerische Standortplanung und regionale
  Wirtschaftsförderung: eine empirische Analyse des Standortverhaltens
  industrieller Grossunternehmen}.
\newblock Vandenhoeck \& Ruprecht, 1983.

\bibitem{maceachren1994visualization}
A.~M. MacEachren.
\newblock Visualization in modern cartography: setting the agenda.
\newblock {\em Visualization in modern cartography}, 28(1):1--12, 1994.

\bibitem{maceachren1995maps}
A.~M. MacEachren.
\newblock {\em How maps work: representation, visualization, and design}.
\newblock Guilford Press, 1995.

\bibitem{maceachren2001research}
A.~M. MacEachren and M.-J. Kraak.
\newblock Research challenges in geovisualization.
\newblock {\em Cartography and geographic information science}, 28(1):3--12,
  2001.

\bibitem{mcmillan1965manufacturers}
T.~McMillan.
\newblock Why manufacturers choose plant locations vs. determinants of plant
  locations.
\newblock {\em Land Economics}, 41(3):239--246, 1965.

\bibitem{munzner2008process}
T.~Munzner.
\newblock Process and pitfalls in writing information visualization research
  papers.
\newblock In A.~Kerren, J.~T. Stasko, J.-D. Fekete, and C.~North, editors, {\em
  Information Visualization: Human-Centered Issues and Perspectives}, pages
  134--153. Springer Berlin Heidelberg, Berlin, Heidelberg, 2008.

\bibitem{peuquet1994s}
D.~J. Peuquet.
\newblock It's about time: A conceptual framework for the representation of
  temporal dynamics in geographic information systems.
\newblock {\em Annals of the Association of american Geographers},
  84(3):441--461, 1994.

\bibitem{roth2013empirically}
R.~E. Roth.
\newblock An empirically-derived taxonomy of interaction primitives for
  interactive cartography and geovisualization.
\newblock {\em IEEE transactions on visualization and computer graphics},
  19(12):2356--2365, 2013.

\bibitem{scheepens2011interactive}
R.~Scheepens, N.~Willems, H.~van~de Wetering, and J.~J. Van~Wijk.
\newblock Interactive visualization of multivariate trajectory data with
  density maps.
\newblock In {\em Pacific Visualization Symposium (PacificVis), 2011 IEEE},
  pages 147--154. IEEE, March 2011.

\bibitem{shaw2009gis}
S.-L. Shaw and H.~Yu.
\newblock A {GIS}-based time-geographic approach of studying individual
  activities and interactions in a hybrid physical--virtual space.
\newblock {\em Journal of Transport Geography}, 17(2):141--149, 2009.

\bibitem{shneiderman1996eyes}
B.~Shneiderman.
\newblock The eyes have it: A task by data type taxonomy for information
  visualizations.
\newblock In {\em Visual Languages, 1996. Proceedings., IEEE Symposium on},
  pages 336--343. IEEE, 1996.

\bibitem{stolte2002polaris}
C.~Stolte, D.~Tang, and P.~Hanrahan.
\newblock Polaris: a system for query, analysis, and visualization of
  multidimensional relational databases.
\newblock {\em IEEE Transactions on Visualization and Computer Graphics},
  8(1):52--65, Jan 2002.

\bibitem{Strotmann2007}
H.~Strotmann.
\newblock Entrepreneurial survival.
\newblock {\em Small business economics}, 28(1):87--104, 2007.

\bibitem{sun2016embedding}
G.~Sun, R.~Liang, H.~Qu, and Y.~Wu.
\newblock Embedding spatio-temporal information into maps by route-zooming.
\newblock {\em IEEE transactions on visualization and computer graphics}, 2016.

\bibitem{tominski2012stacking}
C.~Tominski, H.~Schumann, G.~Andrienko, and N.~Andrienko.
\newblock Stacking-based visualization of trajectory attribute data.
\newblock {\em IEEE Transactions on visualization and Computer Graphics},
  18(12):2565--2574, 2012.

\bibitem{wang2017gaussian}
Z.~Wang, N.~Ferreira, Y.~Wei, A.~S. Bhaskar, and C.~Scheidegger.
\newblock Gaussian cubes: Real-time modeling for visual exploration of large
  multidimensional datasets.
\newblock {\em IEEE Transactions on Visualization and Computer Graphics},
  23(1):681--690, 2017.

\bibitem{weber1909urber}
A.~Weber.
\newblock Urber don standort der industrien.
\newblock {\em JCB Mohr, Tubingen, Germany (Alfred Weber's Theory of the
  Location of Industries: English translation by CJ Friedrich, 1929, University
  of Chicago Press, Chicago)}, 1909.

\bibitem{wehrend1990problem}
S.~Wehrend and C.~Lewis.
\newblock A problem-oriented classification of visualization techniques.
\newblock In {\em Proceedings of the 1st Conference on Visualization '90}, VIS
  '90, pages 139--143, Los Alamitos, CA, USA, 1990. IEEE Computer Society
  Press.

\bibitem{woratschek2000}
H.~Woratschek.
\newblock Standortentscheidungen im handel: M{\"o}glichkeiten und grenzen von
  gravitationsmodellen.
\newblock In {\em Neue Aspekte des Dienstleistungsmarketing}, pages 29--48.
  Springer, 2000.

\bibitem{woratschek2004dienstleistungsmanagement}
H.~Woratschek and S.~Pastowski.
\newblock Dienstleistungsmanagement und standortentscheidungen im
  internationalen kontextm{\"o}glichkeiten und grenzen des einsatzes
  betriebswirtschaftlicher verfahren.
\newblock In {\em Management internationaler Dienstleistungen}, pages 215--240.
  Springer, 2004.

\bibitem{yu2006spatio}
H.~Yu.
\newblock Spatio-temporal gis design for exploring interactions of human
  activities.
\newblock {\em Cartography and Geographic Information Science}, 33(1):3--19,
  2006.

\bibitem{Zelenovic2003}
D.~Zelenovic.
\newblock Location of production systems.
\newblock {\em The Design of Production Systems}, pages 373--394, 2003.

\end{thebibliography}
\end{document}